\pdfoutput=1

\documentclass[english]{article}
\usepackage[T1]{fontenc}
\usepackage[latin9]{inputenc}
\usepackage{color}
\usepackage{array}
\usepackage{amstext}
\usepackage{amssymb}
\usepackage{graphicx}

\makeatletter

\providecommand{\tabularnewline}{\\}

\@ifundefined{date}{}{\date{}}
\usepackage{jinstpub-arxiv}
\usepackage{babel}

\author[a,1]{F. Neves,\note{Corresponding author.}}
\author[a]{A. Lindote,}
\author[a]{A. Morozov,}
\author[a]{V. Solovov,}
\author[a]{C. Silva,}
\author[a]{P. Bras,}
\author[a]{J. P. Rodrigues}
\author[a]{and M. I. Lopes}

\affiliation[a]{LIP-Coimbra, Department of Physics, University of Coimbra, Rua Larga, 3004-516 Coimbra, Portugal}

\emailAdd{neves@coimbra.lip.pt}

\abstract{
The performance of a detector using liquid xenon (LXe) as a scintillator is strongly dependent on the collection efficiency
for xenon scintillation light, which in turn is critically dependent on the reflectance of the surfaces that surround the active
volume. To improve the light collection in such detectors the active volume is usually surrounded by polytetrafluoroethylene
(PTFE) reflector panels, used due to its very high reflectance -- even at the short wavelength of scintillation light of LXe (peaked at $178\,\text{nm}$). 
In this work, which contributed to the overall R\&D effort towards the LUX-ZEPLIN (LZ)
experiment, we present experimental results for the absolute reflectance of three different PTFE samples (including
the material used in the LUX detector) immersed in LXe for its scintillation light. The obtained results show that very high
bi-hemispherical reflectance values ($\geq97\text{\%}$) can be achieved, enabling very low energy
thresholds in liquid xenon scintillator-based detectors.
}


\makeatother

\usepackage{babel}
\begin{document}

\title{Measurement of the absolute reflectance of polytetrafluoroethylene
(PTFE) immersed in liquid xenon}
\maketitle

\section{Introduction\label{sec:Introduction}}

Liquid xenon (LXe) is widely used as sensitive medium in detectors
for different applications ranging from astrophysics to medical imaging
\cite{Aprile:2010aa,Chepel:2013aa}. Most of these detectors are based
on the collection of the xenon scintillation light. This is the case,
for instance, of several detectors used in direct dark matter detection
(e.g. ZEPLIN-II, LUX, XENON, PANDA-X) and other rare event search
experiments (e.g. EXO, MEG).

The performance of these detectors (e.g. their energy resolution,
discrimination capability and sensitivity) strongly depends on the
amount of scintillation light collected, which, in turn, is critically
dependent on the reflectance of the internal surfaces of the detector
active region. Most common construction materials, in particular stainless
steel (SS) and titanium, have very low reflectivity for the xenon
scintillation light ($\text{peak}=178\,\text{nm}$; $\text{FWHM}=14\,\text{nm}$
\cite{Jortner:1965aa}). Hence, the detector active region is usually
surrounded by a good reflector for light in this wavelength region.

Polytetrafluoroethylene (PTFE), a synthetic fluoropolymer, has proved
to be the best choice of reflector for LXe scintillation-based detectors
so far. Besides having a high Vacuum Ultra-Violet (VUV) reflectance
(see below), it presents other very suitable properties: the manufacture
process yields a highly radio-pure material (ppb in U/Th) \cite{Akerib:2015aa,Aprile:2011aa};
it has good mechanical properties (despite a $1.4\text{\%}$ thermal
contraction at liquid xenon temperatures \cite{Kirby:1956aa}); and
outgassing rates are relatively low \cite{POOLE:1980aa}. Unlike some
metallic coatings, such as aluminum, which are also good VUV reflectors,
its optical properties are stable against corrosion (e.g. from atmospheric
$O_{2}$ or $H_{2}O$) therefore not requiring any special handling
or storing conditions \cite{Akerib:2015zz}. Moreover, PTFE is chemically
inert, has a high melting point and excellent dielectric properties
\cite{CHANT:1967aa}.

The reflectance of PTFE has been extensively measured in gas and in
vacuum for wavelengths in the range from $200\,\text{nm}$ up to $3400\,\text{nm}$
\cite{Tsai:2008aa,Weidner:1985aa,Weidner:1981aa}. The results show
a reflectance of about $99\text{\%}$ for wavelengths between $350\,\text{nm}$
and $1500\,\text{nm}$. The reflectance decreases for ultraviolet
light, being $93.4\text{\%}$ at $200\,\text{nm}$. It is also known
that, at least between $250$ and $3400\,\text{nm}$, the optical
properties differ for low- and high-density PTFE \cite{Tsai:2008aa}:
in this wavelength range, low-density PTFE ($1.55\,\text{g/ml}$)
shows a lower transmittance and a higher reflectance compared to the
high-density material ($2.17\,\text{g/ml}$).

Measurements between $120$ and $220\,\text{nm}$ using an Ulbricht
sphere \cite{Kadkhoda:1999aa} show that the reflectance of an unspecified
type of PTFE decreases slowly between $220$ and $175\,\text{nm}$,
followed by a sharp drop at shorter wavelengths due the absorption
edge of the PTFE ($161\,\text{nm}$). At $178\,\text{nm}$, the reflectance
was measured to be about $56\text{\%}$.

More recently, PTFE bi-hemispherical reflectance (BHR) and its angular
distribution for gaseous xenon scintillation light (peaked at $175\,\text{nm}$)
were reported for several samples of PTFE manufactured by different
processes (extruded, expanded, skived and pressed), along with those
of other fluoropolymers, namely poly(ethene-co-tetrafluoroethene)
(ETFE), hexafluoropropylene (FEP) and perfluoroalkoxy (PFA) \cite{Silva:2010aa}.
All the samples were measured in an atmosphere of argon gas at room
temperature. The measurements were carried out using a dedicated angle
resolution system (goniometer), with the reflected light being sampled
at a wide range of angles (including directions outside the plane
of incidence). The obtained BHR for these PTFE samples ranges from
about $47\text{\%}$ to $70\text{\%}$, depending on the manufacturing
process and surface finishing. The best results were obtained with
molded PTFE after polishing. The reflectance distribution of PTFE
at $175\,\text{nm}$ clearly shows specular and diffuse components,
in agreement with what was previously reported for a single angle
of incidence of $45^{\circ}$ at $172\,\text{nm}$ \cite{Bricola:2007aa}.

When installed in LXe detectors, PTFE is immersed in liquid xenon
and kept at low temperature ($\approx-100^{\circ}\text{C}$). Both
these factors are expected to influence the optical properties of
the PTFE surface, making it not trivial to extrapolate the expected
response in these conditions based on the measurements performed in
vacuum or gas \cite{Silva:2009aa}. Attempts of carrying out such
an extrapolation have resulted in predictions which are systematically
and significantly lower than the values $\mathcal{O}(>95\text{\%})$
estimated in LXe detectors using this material as a reflector (such
as ZEPLIN-II, XENON100, and LUX) \cite{Silva:2009aa,Akerib:2012ys}.
PTFE reflectance in excess of $90-95\text{\%}$ has also been reported
for smaller LXe chambers \cite{Yamashita2004aa,neves:2005aa}. In
all these works, a lower limit for PTFE reflectance has been obtained
by comparing the collected scintillation light produced by radioactive
sources, internal or external, in a LXe chamber built for other purposes,
with the expectation obtained by Monte-Carlo (MC) simulation of the
light transport and collection in the chamber. Due to the dimensions
and other features of those chambers, several other parameters apart
from PTFE reflectance are also fitted (or their values fixed based
on assumptions). Examples of those extra parameters are the quantum
efficiency of the photomultipliers, the reflectivity of other materials
existing in the active volume of the chamber (e.g. field grid wires)
and the scintillation light attenuation and Rayleigh scattering lengths
in liquid xenon. Moreover, those simulations have always assumed that
reflections on the PTFE surfaces were purely diffuse, following the
Lambert cosine law \cite{pedrotti:1993aa}.

It is thus highly desirable to measure the absolute value of the reflectance
of PTFE immersed in LXe for its scintillation light in a dedicated
experiment and with a precision better than $O(1\text{\%})$ as a
few percent difference in the PTFE reflectance may have a very significant
impact on the threshold and sensitivity of the detector \cite{Akerib:2015zz}.

In this work, we report results of the absolute value of the reflectance
of several samples of PTFE for xenon scintillation light and with
PTFE directly immersed in LXe. To the best of our knowledge, these
are the first measurements of this kind carried out in a dedicated
experiment and using a method specifically designed for this purpose.
This work was carried out in connection to the R\&D effort towards
the LUX-ZEPLIN (LZ) experiment \cite{Akerib:2015zz}.

\section{Method\label{sec:Method}}

Due to the constraint imposed by the need of having PTFE directly
immersed in LXe, the high purity requirements for the detection of
the xenon scintillation light and the low temperature ($\approx-100^{\circ}\text{C}$),
it is not feasible to use one of the methods commonly employed in
direct reflectance measurements (e.g. methods based on the use of
a Ulbricht sphere or in an angle resolution scattering system). Also,
one must keep in mind that a direct measurement of the reflectance
using those methods is actually not possible and that detailed MC
simulations of light propagation in the liquid are always necessary
to account for Rayleigh scattering and light absorption in LXe, with
the latter depending on the experimental conditions, namely, the cleanliness
and outgassing from detector construction materials.

The method presented in this work is based on the measurement of the
relative amount of light, produced directly in LXe by short-range
mono-energetic alpha particles, that is collected at a photodetector
after traversing a chamber where the geometry of the optical surfaces/volumes
(PTFE/LXe) can be varied \textit{on the fly} with high precision.
This key feature allowed to measure light collection for different
geometries, so varying the average number of reflections at the PTFE
from just a couple to a few tens. Similarly, the average track length
for optical photons in the liquid spans from a few centimeters to
a couple of meters, depending on the liquid purity. The experimental
results were then fitted with those from a detailed MC simulation
of the light transport for each of the considered geometries, using
the optical properties of PTFE and absorption length in the LXe as
free parameters. Both the simulation and fitting procedures are explained
in more detail in sec.~\ref{sec:Data-analysis-and-results}.

The chamber was built using only materials compliant with the purity
requirements of LXe. In the active region, materials other than PTFE
were kept to a bare minimum to reduce additional uncertainties from
modeling their reflectances.

\section{Experimental setup\label{sec:Experimental-setup}}

\begin{figure}
\begin{centering}
\includegraphics[height=0.3\textheight]{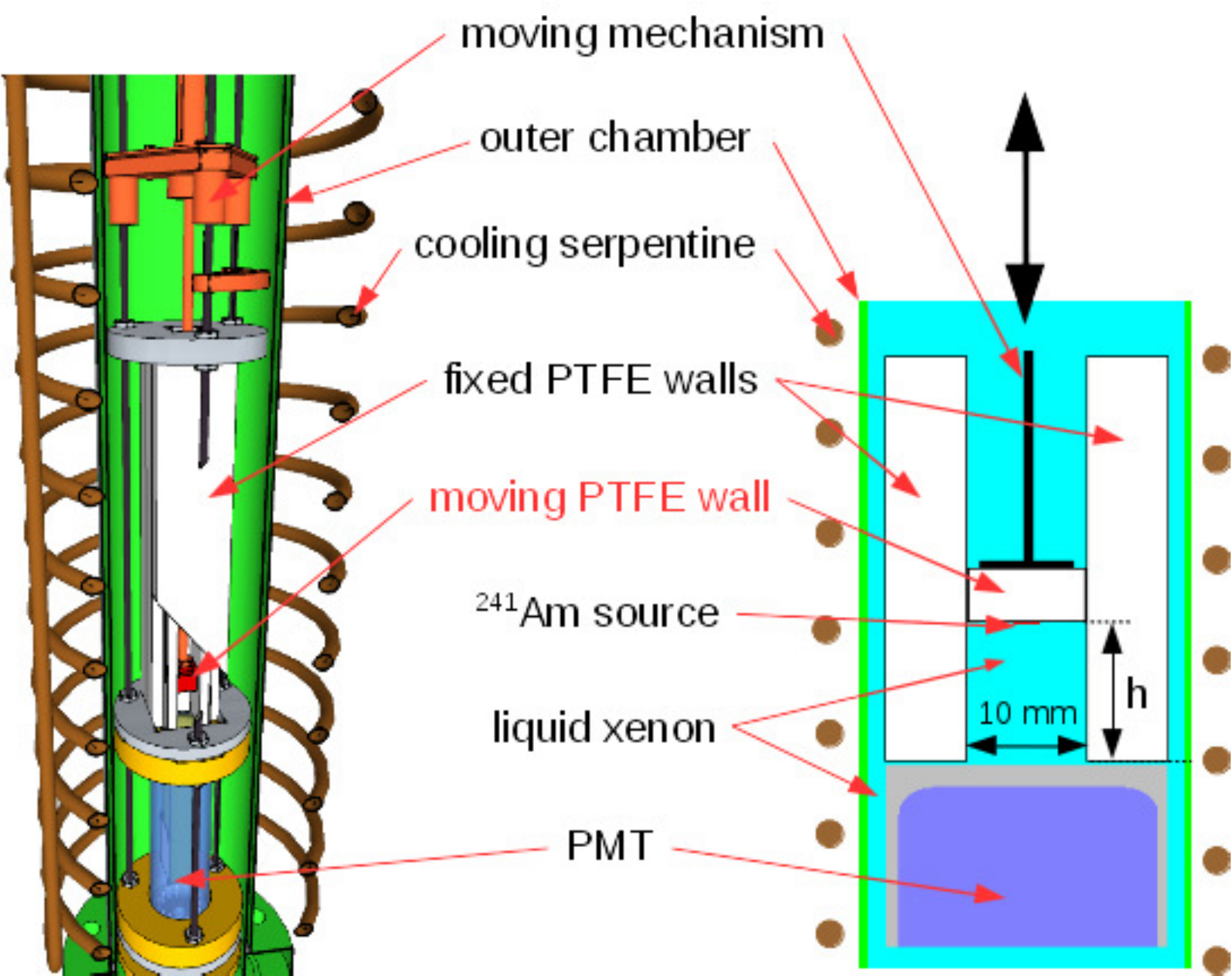}
\par\end{centering}
\caption{\label{fig:setup}3D view (left) and schematic representation (right)
of the chamber showing the PTFE sample walls, the $^{241}Am$ source
placed under the moving wall and the PMT at the bottom.}
\end{figure}

The set-up used for the measurements reported here is schematically
represented in fig.~\ref{fig:setup}. Both the lateral and the top
walls of the active region are made from $1\,\text{cm}$ thick PTFE
whose reflectance is intended to be measured. The lateral walls are
arranged to form a cavity $150\,\text{mm}$ long and with a square
section of $10\times10\,\text{mm}^{2}$. The top wall also has a square
section of $10\times10\,\text{mm}^{2}$ and can be moved by means
of a micrometer drive along the length of the cavity, thus changing
the chamber geometry through its height $h$ (see fig.~\ref{fig:setup}).
Centered on the top wall and recessed by $0.1\,\text{mm}$ there is
a $^{241}Am$ source ($6.8\,\text{mm}$ in diameter) deposited on
a stainless steel (SS) plate with a diameter of $8.46\,\text{mm}$.
The $^{241}Am$ source emits alpha particles with an energy of $5.486\,\text{MeV}$
at a rate of $\approx1\,\text{kBq}$. Because these particles have
a short range in LXe ($\approx50\,\mu\text{m}$) compared with the
dimensions of the active region of the chamber, we henceforth consider
the energy of the alpha particle to be fully deposited in a point-like
interaction. The energy deposited by each alpha particle in the LXe
is then converted into $\approx3.36\times10^{5}$ VUV photons from
xenon scintillation ($\text{peak}=178\,\text{nm}$; $FWHM=14\,\text{nm}$)
\cite{Miyajima:1992aa,Jortner:1965aa}. Some of those scintillation
photons are absorbed by impurities in the LXe bulk volume or at any
of the interfaces delimiting the active region (PTFE or SS from the
source). The remaining photons, after traveling through the LXe volume
and being reflected at the various surfaces, are detected by a photomultiplier
(Hamamatsu R1668) placed at the bottom of the PTFE cavity and facing
the $^{241}Am$ source. The PMT window has a diameter of $28.6\,\text{mm}$
and is made of UV grade fused silica with a transmission of $\approx90\text{\%}$
for xenon scintillation light. There is a $1\,\text{mm}$ gap between
the PTFE walls and the PMT window to allow the displacement of LXe
in and out from the cavity when moving the top wall and filling or
emptying the chamber.

The described apparatus (fig.~\ref{fig:setup}) is placed inside
a vacuum tight cylindrical vessel, made of SS, connected to the xenon
gas handling and purification system. All parts of the chamber, with
the exception of the PMT, were cleaned in an ultrasonic bath of pure
acetone prior to assembly. After closing, the chamber was heated up
to $\approx50{}^{\circ}\text{C}$ and pumped until a vacuum of $<5\times10^{-8}\,\text{mbar}$
was achieved. This was followed by circulating xenon gas through the
chamber and an Oxisorb purifier for at least $5$ days. To control
the purity of the xenon, the electron lifetime in the liquid phase
was measured in a separate parallel plate chamber also connected to
the gas system. The electron lifetime in the liquid is extracted directly
from the pulse shape analysis of the charge signals due to a $^{207}$Bi
source placed at the cathode plate \cite{Aprile:1991aa}. For all
the reported measurements the electron lifetime was better than $40\,\text{\ensuremath{\mu}s}$
(at $1\,\text{kV/cm}$), the highest value that can be measured with
that chamber. This value allows to estimate the oxygen-equivalent
concentration of electro-negative impurities in the LXe to be less
than $17\,\text{ppb}$ \cite{Bakale:1976aa}. Although being used
as a control parameter of the LXe purity across the various measurements,
in fact this value cannot be used to calculate the absorption length
for the xenon scintillation light as will be discussed in sec.~\ref{subsec:Results}.

The setup was cooled down to $-103\pm1^{\circ}\text{C}$ by placing
the chamber in a cooled ethylene bath. The temperature of this bath
is controlled by the amount of liquid nitrogen circulating through
a copper coil surrounding the outer vessel (fig.~\ref{fig:setup})
and an heater placed at the bottom of the bath. The chamber is filled
with $\sim1.5\,\text{bar}$ of xenon and left to thermalize for $\sim12\,\text{hours}$
prior to start condensing xenon. Xenon is then slowly condensed into
the chamber until it completely covers the PTFE walls. During condensation
the position of the liquid level is monitored using a capacitive level
sensor placed inside the outer chamber but outside the PTFE walls.
The level of the liquid along with its temperature and the pressure
in the gas phase are continuously monitored and stored for future
reference.

\begin{figure}
\begin{centering}
\includegraphics[width=0.46\textwidth,height=0.2\textheight]{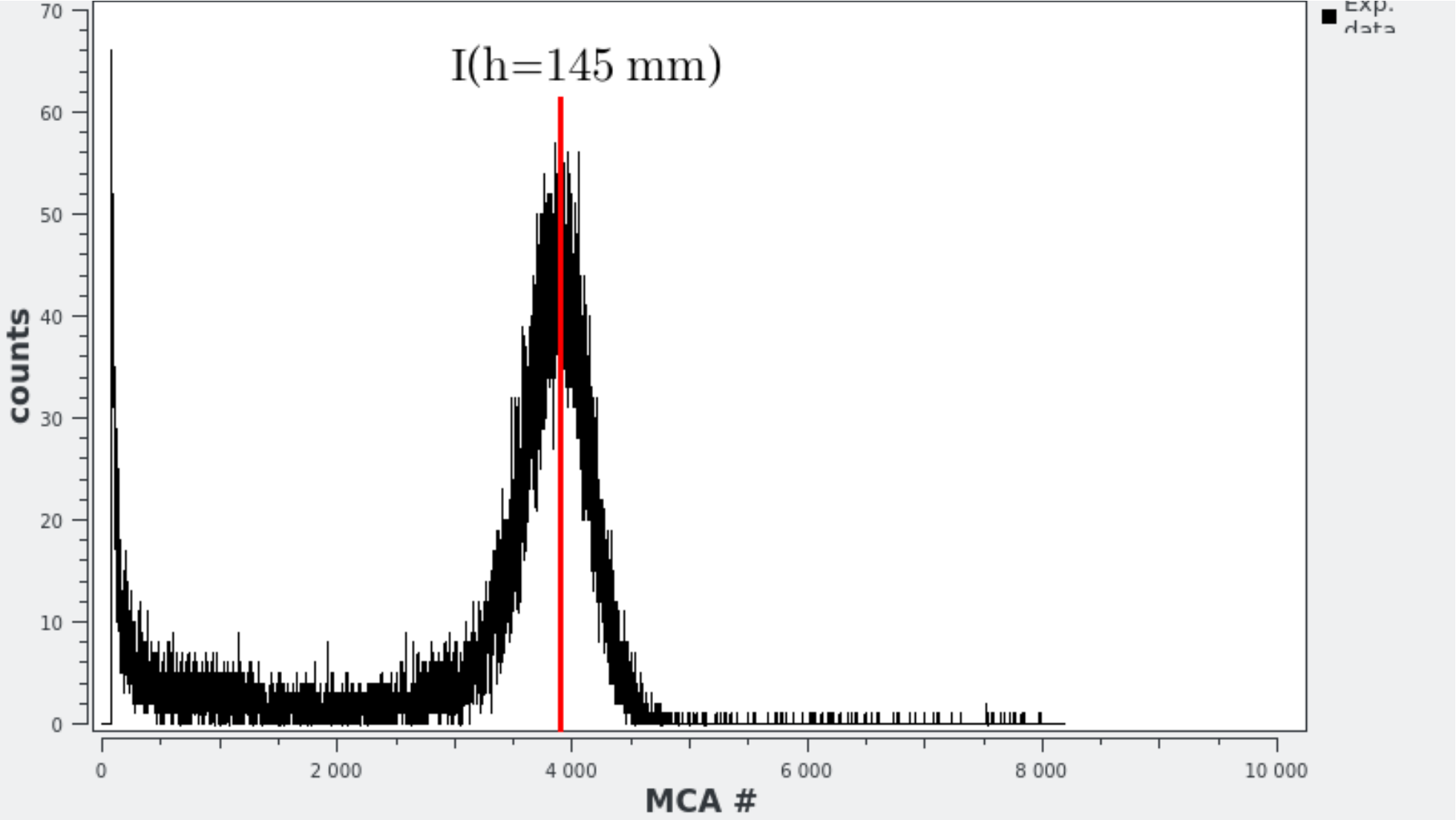}\includegraphics[width=0.46\textwidth,height=0.2\textheight]{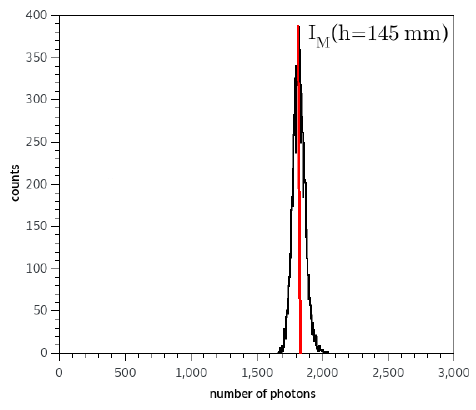}
\par\end{centering}
\caption{\label{fig:spectrum}Example of an experimental spectrum obtained
using one of the PTFE samples with the source at $h=145\,\text{mm}$
(left) ; Example of a simulated spectrum of the number of photons
crossing the PMT window also for $h=145\,\text{mm}$ (right). The
values of $I(145)$ and $I_{\mathcal{M}}(145)$ (where $\mathcal{M}$
tags the optical model used to simulate the PTFE reflectance - sec.~\ref{subsec:Data-analysis})
are also represented and were obtained by fitting a Gaussian to the
correspondent peak in the spectrum.}
\end{figure}

For each position of the top wall ($h$) the light spectrum is acquired
and the position of the peak, $I(h)$, corresponding to the full energy
deposition from the alpha particles in the liquid is estimated from
a Gaussian fit (fig.~\ref{fig:spectrum}). For each PTFE sample the
values of $I$ were obtained for $h$ between $19$ and $145\,\text{mm}$
in steps of $7\,\text{mm}$, both while moving down the top wall closer
to the PMT window and then back up to the farthest position. This
scan was repeated to look for any systematic effects, for example,
from liquid purity degradation with time or bubble formation at the
lateral PTFE walls. To test for reproducibility, the above scanning
procedure was also repeated during successive cooldown cycles separated
by a complete warm up of the chamber. It was observed that for each
PTFE sample the values of $I(h)$ from multiple scans fully agree
within the statistical errors and were therefore merged into a single
dataset for subsequent analysis.

It was observed that the variation of the temperature of the liquid
xenon induced a relative variation in the response of the PMT of $(0.013\pm0.001){}^{\circ}\text{C}{}^{-1}$.
This is attributed to the fact that both the PMT and its voltage divider
are immersed directly in the liquid. All data points presented in
this work were therefore corrected for temperature variations and
normalized to a liquid temperature of $-103^{\circ}\text{C}$.

The PMT anode signals are pre-amplified using a Canberra 2005 module
placed outside the chamber (at room temperature) and then fed into
a Canberra 2020 amplifier for shaping and further amplification. The
shaped signals are digitalized using a multi-channel analyzer (PTG
8008) and the corresponding spectra stored for offline analysis.

\section{Data analysis and results\label{sec:Data-analysis-and-results}}

\subsection{Simulation\label{subsec:Simulation}}

As outlined in sec.~\ref{sec:Method}, to determine the reflectance
from $I(h)$ (sec.~\ref{sec:Experimental-setup}) the variation of
the light collection with the chamber geometry $h$ has to be modeled
by MC simulation. For that, a detailed simulation of the transport
of the scintillation photons for each of the considered geometries
was implemented using the ANTS2 software toolkit \cite{Morozov:2016aa}.
Only the inner chamber was simulated, as the details of the outer
cylindrical vessel, structural support and instrumentation elements
are irrelevant for light propagation inside the inner cell. The results
from the ANTS2 simulations were validated at an early stage against
those obtained using the GEANT4 toolkit \cite{Agostinelli:2003aa}
for the exact same conditions. As the results obtained using both
packages agree within statistical errors, ANTS2 was selected for the
rest of this study due to the considerably shorter computational time
required for the simulation of the propagation of such a large number
of initial scintillation photons.

For the simulation of the light transport in the inner cell, the relevant
physical processes to be modeled are the propagation of the optical
photons in the bulk LXe volume and their behavior on the interfaces.
For the propagation of light in xenon, Rayleigh scattering and bulk
absorption in LXe were considered: the interaction length for the
former was set to $29\,\text{cm}$ \cite{Seidel:2002vf}, while the
latter (which is strongly affected by xenon purity) was left as a
free parameter and allowed to vary during the analysis stage. Given
that most of the inner cavity is surrounded by the PTFE for which
we intend to measure the reflectance, the large majority of the interactions
of the optical photons are with this material. For the simulation
of the behavior of photons in the xenon/PTFE interfaces two models
were considered. The first model ($D$) considers only diffuse reflection
following the Lambert cosine law 

\begin{equation}
dR_{dif}(\theta_{r})=Acos(\theta_{r})d\theta_{r}\text{,}\label{eq:lambert}
\end{equation}
where $R_{dif}(\theta_{r})$ represents the probability of the light
being reflected back into the LXe with an angle $\theta_{r}$ relative
to the PTFE surface normal. The free parameter $A$ is the albedo,
a constant representing the integral probability of the light not
being absorbed at the PTFE. The second model ($DS$) considers specular
reflection at the liquid/PTFE interface and diffuse reflection of
the photons that are refracted into the PTFE. To calculate the probability
of specular reflection at any given angle of incidence ($\theta_{i}$)
the Fresnel equations were considered for non-polarized light:

\begin{equation}
R_{spec}(\theta_{i})=\frac{1}{2}\left(R_{p}(\theta_{i})+R_{s}(\theta_{i})\right)\text{,}\label{R_spec}
\end{equation}
where $R_{p}(\theta_{i})$ and $R_{s}(\theta_{i})$ represent, respectively,
the probability of reflection for p- and s-polarized light \cite{pedrotti:1993aa}:

\begin{equation}
R_{p}(\theta_{i})=\left(\frac{n^{2}cos\theta_{i}-\sqrt{n^{2}-sin^{2}\theta_{i}}}{n^{2}cos\theta_{i}+\sqrt{n^{2}-sin^{2}\theta_{i}}}\right)^{2}
\end{equation}
and

\begin{equation}
R_{s}(\theta_{i})=\left(\frac{cos\theta_{i}-\sqrt{n^{2}-sin^{2}\theta_{i}}}{cos\theta_{i}+\sqrt{n^{2}-sin^{2}\theta_{i}}}\right)^{2}\text{,}
\end{equation}
with $n=n_{PTFE}/n_{LXe}$, where $n_{PTFE}$ and $n_{LXe}$ correspond
to the refractive index of PTFE and liquid xenon, respectively. The
value of $n_{LXe}$was considered to be constant and set to $1.69$
\cite{Solovov:2003ax} while $n_{PTFE}$ was left as a free parameter.
The probability of diffuse reflection can then be expressed as
\begin{equation}
R_{dif}(\theta_{i})=A\left[1-R_{spec}(\theta_{i})\right]\text{,}\label{eq:R_dif}
\end{equation}
where $1-R_{spec}(\theta_{i})$ is the probability of light being
refracted into the PTFE bulk volume. The direction of this light when
exiting the PTFE also follows the Lambert cosine law as for the $D$
model (eq.~\ref{eq:lambert})

\begin{equation}
dR_{dif}(\theta_{r}|\theta_{i})=R_{dif}(\theta_{i})cos(\theta_{r})d\theta_{r}\text{.}\label{eq:dR_dif}
\end{equation}
It is worth noting that the $DS$ model coincides with the $D$ model
when $n_{PTFE}=n_{LXe}$ corresponding to $R_{spect}=0$. In this
particular case eq.~\ref{eq:R_dif} can be written as $R_{dif}=A$,
having no dependence on $\theta_{i}$, $n_{PTFE}$ or $n_{LXe}$.

\begin{figure}
\begin{centering}
\includegraphics[width=0.46\textwidth,height=0.2\textheight]{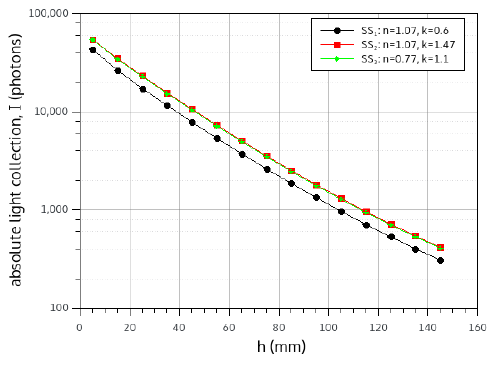}\includegraphics[width=0.46\textwidth,height=0.2\textheight]{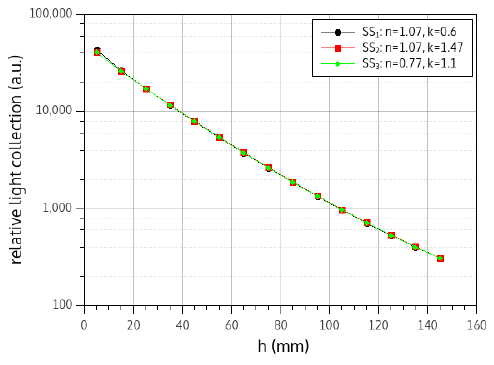}
\par\end{centering}
\caption{\label{fig:SS}Example of the simulated absolute light collection
as a function of the chamber height ($h)$ for $A=0.95$, $n_{PTFE}=1.75$
($DS$ model) and three different SS refractive indexes of the $^{241}\text{Am}$
source (left). Despite the significant different refractive indexes
and correspondent values of the absolute light collection, the scaled
light collection profiles match perfectly within the statistical errors
(right).}
\end{figure}

Apart from PTFE, the only materials where light can be reflected inside
the inner cell are the SS in the $^{241}Am$ source and the fused
silica in the PMT window. Despite of contributing only with a very
small fraction to the total internal area of the chamber, the SS surface
is the first interface for $\approx50\text{\%}$ of the photons, that
are produced just above the source along the very short range ($\approx50\,\mu\text{m}$)
alpha particle tracks. As the composition of this SS is unknown, we
tested two pairs of refractive index values ($(n,k)=(0.77,1.1)$ and
$(n,k)=(1.07,1.47)$) found in the literature \cite{Karlsson:1982aa}
(corresponding to two samples of SS with different compositions),
and an additional third value ($(n,k)=(1.07,0.6)$) which was estimated
using data available from a previous experiment performed by our group
for yet a different SS \cite{Solovov:2003ax}. Despite the large differences
in total reflectivity between the tested refractive indexes ($10\%-30\text{\%}$
for normal incident light), leading to a very different number of
detected photons for every $h$ (fig.~\ref{fig:setup}), it was verified
that the scaled light collection profiles match perfectly within statistical
errors (fig.~\ref{fig:SS}). The best match between any two of the
light collection profiles (for different SS refractive indexes) was
obtained by multiplying one of them by a scaling factor $C$ and minimizing
the difference between the scaled and unscaled profiles using $C$
as a free parameter. As not affecting the relative light collection
(see sec.~\ref{subsec:Data-analysis}), the result of all simulations
presented in this work were done considering a refractive index for
the SS of $n=1.07$, with a complex factor $k=0.6$ accounting for
the attenuation at the SS.

Regarding the PMT, to our best knowledge, its quantum efficiency (QE)
at LXe temperature is somewhere in the range of $15\%-30\text{\%}$
for the xenon scintillation light. As for the case of the SS reflectivity,
different values of QE result in a simple scaling in the number of
detected photons as a function of $h$. For that reason all the photons
arriving at the PMT window and refracted inside are considered to
be detected with $100\text{\%}$ efficiency, thus also disregarding
any loss due to the window thickness as well as any inefficiency in
the collection of the the photoelectrons on the first dynode. The
light reflection at the PMT window was considered to be purely specular
with an index of refraction of the fused silica equal to $1.59$ \cite{Gupta:1998aa}.

\textcolor{red}{}

\subsection{Data analysis\label{subsec:Data-analysis}}

The PTFE reflectance is obtained by fitting the measured relative
light collection, $I(h)$, with the prediction of the Monte Carlo
simulation. As detailed in the previous section, the uncertainties
in both the reflectivity of the SS and QE of the PMT do not allow
for a direct comparison of the experimental and simulated results
in terms of the absolute number of photons arriving at the PMT photocathode.
Instead, we compare the relative variation of the light collected
with $h$ assuming that both the SS reflectivity and PMT QE contributions
correspond to a scaling factor.

For the simplest model considering only diffuse reflection in the
PTFE ($D$), the free fitting parameters were the PTFE albedo ($A$,
eq.~\ref{eq:lambert}) and the attenuation length ($\lambda$) for
the scintillation light in LXe. When considering specular reflection
($DS$), the refractive index of the PTFE ($n_{PTFE}$, eq.~\ref{R_spec})
is an additional free parameter. For each source position ($h$) and
set of parameters $\mathcal{O}_{D}=\left\{ A,\lambda\right\} $ or
$\mathcal{O}_{DS}=\left\{ A,n_{PTFE},\lambda\right\} $, depending
on the model, we sample the isotropic emission of $10^{5}$ scintillation
photons per alpha particle for a total of $300$ alpha particles distributed
uniformly over the $^{241}Am$ source surface. The simulated number
of scintillation photons per alpha particle is $\approx1/3$ of the
real number of photons one expects to be produced in the liquid (sec.~\ref{sec:Experimental-setup}).
Since we are only interested in the relative variation of the light
collection and to save computational resources, we reduced the number
of emitted photons in the simulation while keeping the statistical
errors on $I_{\mathcal{M}}$ lower than $\mathcal{O}(0.1\%)$ ($I_{\mathcal{M}}(h)\gtrsim1000$),
where $\mathcal{M}$ stands for the $D$ or $DS$ model. Given the
CPU time required per simulation, we chose to generate a multidimensional
template grid for each of the reflectivity models. This approach also
allows for a more systematic search of local minima and possible degeneracy
of solutions. For any given set of parameters $\mathcal{O_{M}}$,
the simulated response $I_{\mathcal{M}}(\mathcal{O_{M}};h)$ is obtained
using a bilinear interpolation at the correspondent grid nodes. Henceforth
and for the sake of simplicity, the subscript $\mathcal{M}$ will
be omitted from the set of optical parameters $\mathcal{O_{M}}$ when
the model is already defined by the calling function (e.g. $I_{\mathcal{M}}(\mathcal{O_{M}};h)\equiv I_{\mathcal{M}}(\mathcal{O};h)$).

For the diffuse only model ($D$) a $39\times50\times29$ grid was
generated ($56550$ individual simulations) within the corresponding
ranges: $A=[0.6,1]$, $\lambda=[100,5000]\,\text{mm}$ and $h=[5,145]\,\text{mm}$.
Both the $\lambda$ and $h$ axis are regular with a step size of
$100\,\text{mm}$ and $5\,\text{mm}$, respectively. For the $A$
axis, we refined the step size from $0.05$ for $A\le0.7$ to $0.01$
in the range $0.7<A<0.9$ and $0.005$ for $A\ge0.9$, corresponding
to the region of interest for all the samples measured in this work.
For the model considering specular reflection ($DS$) a $16\times29\times29\times29$
grid was generated (corresponding to a total of $390224$ simulations)
with the same ranges defined above for the $A$, $\lambda$ and $h$
axis. The refractive index ($n_{PTFE}$) is varied in the range $[1.3,1.9]$
with a step size of $0.01$ for $1.65\leq n_{PTFE}\leq1.9$ and $0.05$
below $1.65$. Given the quite large number of simulations and corresponding
CPU time required to populate the 4-dimensional grid for the $DS$
model, the generated grid is sparser than for the former case in all
parameters but $h$: the step size for the $\lambda$ axis was kept
at $100\,\text{mm}$ only for $\lambda<1200\,\text{mm}$, where the
variation of the light collected is steeper, and increased to $200\,\text{mm}$
for higher attenuation lengths. Similarly, the albedo axis step size
was increased by a factor of $2$ when compared with the $D$ model,
being $0.01$ in the region of interest, i.e. $A\geqslant0.9$. As
a general rule to refine the grid, the interpolation errors were kept
within the level of the statistical errors coming from the simulation
on the number of photons arriving at the PMT window.

For a given set of experimental points corresponding to a PTFE sample,
the relative variation of the light collected with $h$ is compared
with the simulated response ($I_{\mathcal{M}}$) after scaling the
latter by a constant value $C$. Explicitly, the value of $C$ is
obtained by minimizing

\begin{equation}
\chi_{\mathcal{M}}^{2}(C;\mathcal{O})=\frac{1}{N-N_{\mathcal{O}}-1}\sum_{i}\frac{\left[I(h_{i})-C\times I_{\mathcal{M}}(\mathcal{O};h_{i})\right]^{2}}{\sigma_{exp}^{2}(h_{i})+\sigma_{\mathcal{M}}^{2}(\mathcal{O};h_{i})+\mathcal{\mathcal{H}}_{\mathcal{M}}^{2}(\mathcal{O};h_{i})}\label{eq:Xi2}
\end{equation}
where $i$ represents a data point measured at a given source position
$h_{i}$, $N$ is the total number of experimental points in the data
set and $N_{\mathcal{O}}$ the number of free parameters for the model
being fitted ($N_{D}=2$, $N_{DS}=3$). The best set of optical parameters
is then found by simply iterating through the grid space and finding
the set of parameters $\mathcal{O_{M}}$ yielding the smaller value
for $\chi_{\mathcal{M}}(\mathcal{O})$ (eq.~\ref{eq:Xi2}). The size
of the iteration steps is defined over the region of interest based
on the variation of $\chi_{\mathcal{M}}$ with $\mathcal{O_{M}}$,
and set typically to be a factor of $\lesssim0.1$ smaller than the
$1$ sigma region in $\chi_{\mathcal{M}}$ around the absolute minimum.
The statistical experimental errors, $\sigma_{exp}(h_{i})$ are quadratically
added to the statistical errors from the simulation, $\sigma_{\mathcal{M}}(\mathcal{O};h_{i})$,
and to $\mathcal{H}_{\mathcal{M}}(\mathcal{O};h_{i})$, which takes
into account the uncertainties in the determination of the source
position. In order to propagate the uncertainties in $h$ ($\sigma_{h}$),
we consider the quadratic variation of $I_{\mathcal{M}}$ at the limits
of the interval $\left[h-\sigma_{h},h+\sigma_{h}\right]$ to be representative
of the uncertainties in the light collection from the determination
of $h$, i.e.:

\begin{equation}
\mathcal{\mathcal{H}}_{\mathcal{M}}^{2}(\mathcal{O};h)=(I_{\mathcal{M}}(\mathcal{O};h-\sigma_{h})-I_{\mathcal{M}}(\mathcal{O};h))^{2}+(I_{\mathcal{M}}(\mathcal{O};h+\sigma_{h})-I_{\mathcal{M}}(\mathcal{O};h))^{2}
\end{equation}

It was verified, with the moving mechanism exposed inside a clean
room, that the uncertainty from repositioning the source at any given
position between $h=0\,\text{mm}$ and $h=150\,\text{mm}$ was $\sigma_{h}\lesssim0.1\,\text{mm}$
when measuring its position relatively to the two stoppers placed
at both ends of the allowed course of movement.

\subsection{Results\label{subsec:Results}}

The reflectance of three samples of PTFE was measured: $807NX$ and
$NXT85$ from Applied Plastics Technology (APT) and $8764$ from Technetics.
To the best of our knowledge there is no published data concerning
the reflectance of the first two sample materials for the xenon scintillation
light at $-100^{\circ}\text{C}$ and when immersed in LXe. The 3rd
sample corresponds to the PTFE used in the LUX experiment to define
the active region, thus allowing a direct comparison with the reflectance
estimated b the LUX collaboration \cite{Akerib:2012ys,Szydagis:2014aa}.

\begin{figure}
\begin{centering}
\includegraphics[width=0.6\textwidth]{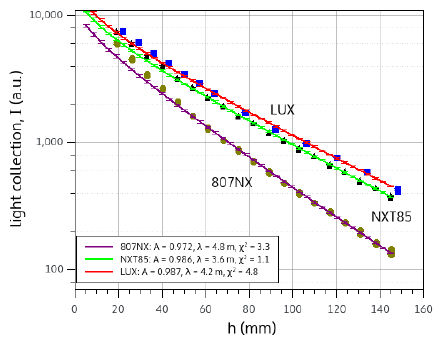}
\par\end{centering}
\caption{\label{fig:Fit-lambert} Fit (purple, green and red lines) to the
experimental data (dark yellow circles, black triangles and blue squares)
using the diffuse-only model ($D$) for the three samples of PTFE
measured, respectively: 807NX, NXT85 and 8764 (LUX). The best fitting
parameters and corresponding $\chi_{D}^{2}$ (eq.~\ref{eq:Xi2})
are also indicated for each sample.}
\end{figure}

Figure.~\ref{fig:Fit-lambert}, shows the fitting results for the
three measured PTFE samples using the diffuse only ($D$) model. It
can be observed that this model systemically underestimates the light
detected for $h\lesssim40\,\text{mm}$, which is attributed to the
absence of specular reflection in the $D$ model. In fact, for small
values of $h$ the average number of reflections decreases and thus
a single specular reflection can provide a direct path to the PMT
window. On the other hand, as the average number of reflections increases
with $h$, the diffuse reflection (which is characterized by large
values of $A$ for the samples studied) becomes dominant thus suppressing
the weight of possible forward specular-only trajectories.

\begin{figure}
\begin{centering}
\includegraphics[width=0.6\textwidth]{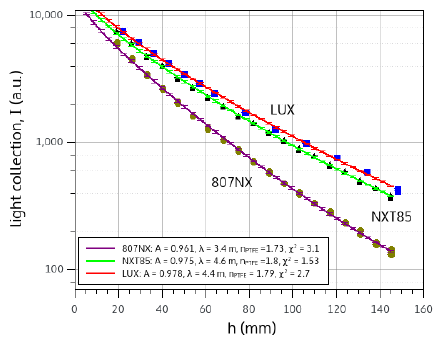}
\par\end{centering}
\caption{\label{fig:Fit-Lamb+Fresn}Fit (purple, green and red lines) to the
experimental data (dark yellow circles, black triangles and blue squares)
using the diffuse plus specular ($DS$) reflection model for the three
samples of PTFE measured, respectively: 807NX, NXT85 and 8764 (LUX).
The best fitting parameters and corresponding $\chi_{DS}^{2}$ (eq.~\ref{eq:Xi2})
are also indicated for each sample.}
\end{figure}

The above interpretation is strengthened when fitting the experimental
results with the model which takes into consideration the specular
reflection at the PTFE walls. As can be seen in fig.~\ref{fig:Fit-Lamb+Fresn},
the $DS$ model succeeds to reproduce the measured relative light
collection for all three PTFE samples.

The results obtained for all the PTFE samples using both reflectivity
models considered in this work are summarized in tab.~\ref{tab:results}.
It is worth noting that, as it can be seen in figs.~\ref{fig:Fit-lambert}
and \ref{fig:Fit-Lamb+Fresn}, the absolute values of the light collected,
$I(h)$ (expressed in a.u.), are scaled accordingly to the corresponding
PTFE reflectivities (tab.~\ref{tab:results}) for all heights and
especially for higher values of $h$, where the larger number of reflections
strongly enhance the effect of small differences in the reflectance.

\begin{table}
\begin{centering}
\begin{tabular}{|c|>{\centering}p{0.1\textwidth}|>{\centering}p{0.1\textwidth}|>{\centering}p{0.1\textwidth}|>{\centering}m{0.1\textwidth}|>{\centering}m{0.1\textwidth}|>{\centering}p{0.1\textwidth}|}
\cline{2-7} 
\multicolumn{1}{c|}{} & \multicolumn{2}{c|}{Diffuse model ($D$)} & \multicolumn{4}{c|}{Diffuse + Specular model ($DS$)}\tabularnewline
\cline{2-7} 
\multicolumn{1}{c|}{} & \multicolumn{1}{>{\centering}p{0.1\textwidth}|}{$A$} & $\lambda\,\text{(mm)}$ & $A$ & $n_{PTFE}$ & $\lambda\,\text{(mm)}$ & $BHR$\tabularnewline
\hline 
807NX & $0.972$

($>0.97$) & $4800$ & $0.961$

($>0.955$) & $1.73$ & $4600$ & $0.961$

($>0.955$)\tabularnewline
\hline 
NXT85 & $0.986$

($>0.984$) & $3600$ & $0.975$

($>0.973$) & $1.8$ & $4600$ & $0.975$

($>0.973$)\tabularnewline
\hline 
LUX & $0.987$

($>0.985$) & $4200$ & $0.978$ ($>0.975$) & $1.79$ & $3000$ & $0.978$

($>0.975$)\tabularnewline
\hline 
\end{tabular}
\par\end{centering}
\caption{\label{tab:results} Best fitting parameters to the experimental data
for the three PTFE samples measured in this work using both the $D$
and $DS$ model for the PTFE reflectance. For the albedo ($A$) the
lower $95\text{\%}$ confidence boundary is also shown. For the $DS$
model the values of the Bi-Hemispherical Reflectance (BHR) are represented
for the best fit parameters and at the lowest $95\text{\%}$ confidence
boundary. Note that for the $D$ case the $BHR$ equals $A$ by definition
of the model. For the $DS$ model, the presented values of the BHR
also equals $A$ for all the studied PTFE samples and within the precision
of the measurement. However, in this case this is just a consequence
of the relative importance of both components of reflection (diffuse/specular)
in the BHR computation which gives little weight to grazing angles
where the specular reflection is more important.}
\end{table}

\begin{figure}
\begin{centering}
\includegraphics[width=0.6\textwidth]{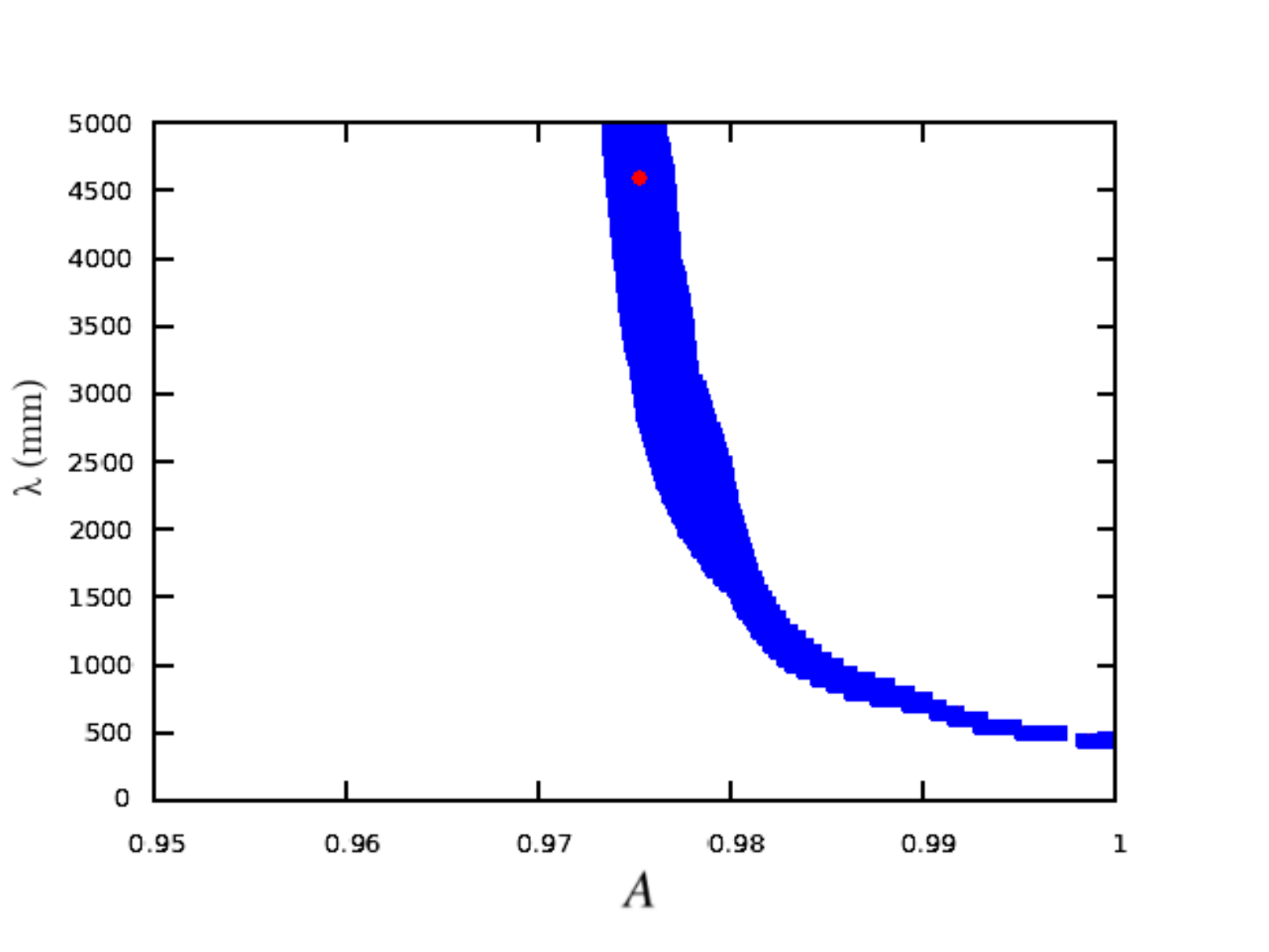}
\par\end{centering}
\caption{\label{fig:A_vs_Lambda_Xi2}Best fit using the $DS$ model of the
NXT85 experimental data (red dot); and, cut ($n_{PTFE}=1.8$) across
the $\lambda$ and $A$ parameters space of the corresponding $95\%$
confidence region (blue band).}
\end{figure}

\begin{figure}
\begin{centering}
\includegraphics[width=0.46\textwidth,height=0.2\textheight]{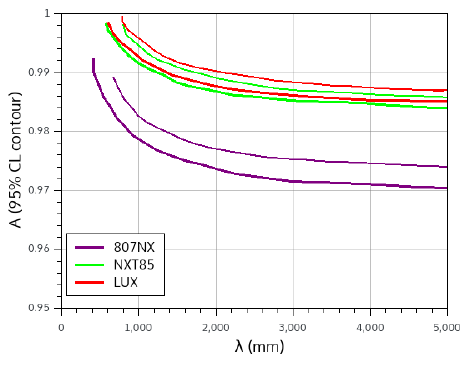}\includegraphics[width=0.46\textwidth,height=0.2\textheight]{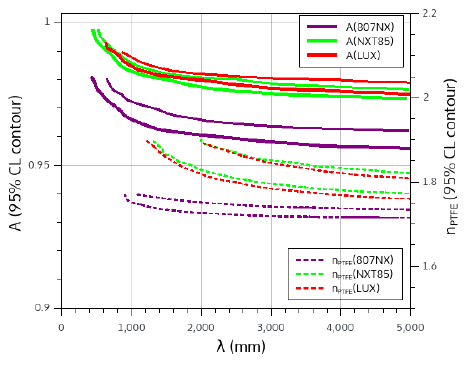}
\par\end{centering}
\caption{\label{fig:95perCL}Upper and lower boundaries of the $95\text{\%}$
confidence regions for the fitting parameters $A$ (solid lines) and
$n_{PTFE}$ (broken lines) as a function of $\lambda$ for both the
$D$ (left) and $DS$ (right) models.}
\end{figure}

Figure~\ref{fig:A_vs_Lambda_Xi2} shows a cut of the $95\text{\%}$
confidence level region from fitting the NXT85 experimental data using
the $DS$ model and illustrates the existence of a degeneracy between
the values of $\lambda$ and $A$, quantities characterizing the extinction
of photons in the liquid and the PTFE, respectively. The profile of
this degeneracy is common to all PTFE samples and just exposes the
inability to resolve which process is actually responsible for the
absorption of the photons. Although the lifetime of free electrons
(sec.~\ref{sec:Experimental-setup}) is a proxy for the concentration
of electronegative impurities in the liquid, there is no direct way
to use this information to independently estimate the actual attenuation
length ($\lambda$) since we do not know which species are actually
diluted in the LXe. 

Given the impossibility to further constrain the attenuation length
in the liquid, the uncertainties on the fitting parameters defining
the reflectance of the PTFE ($A$ and $n_{PTFE}$) were quantified
by the boundaries of their $95\text{\%}$ confidence regions over
the entire range of values of $\lambda$ considered in this work.
The upper and lower boundaries of these regions are represented in
fig.~\ref{fig:95perCL} for the albedo ($A$) and refractive index
of the PTFE ($n_{PTFE}$) for the $D$ and $DS$ models. The lowest
values of the lower $95\text{\%}$ confidence boundary for the albedo
($A_{min}$) are also summarized in tab.~\ref{tab:results}. For
all measured PTFEs samples, $A_{min}$ values correspond to $\lambda=5\,\text{m}$,
as the boundaries asymptote for higher values of the attenuation length.
The values of the Bi-Hemispherical Reflectance (BHR) for the $DS$
model corresponding to both the best fit parameters and the $95\text{\%}$
CL lower limits are also shown in tab.~\ref{tab:results} (the latter
in parenthesis). The BHR values are obtained for a white sky illumination
\cite{nicodemus:1977aa} and are dominated by the correspondent value
of $A$, being almost insensitive to $n_{PTFE}$ within its $95\text{\%}$
confidence region at any attenuation length. The reason for this is
that the incident grazing angles for which $R_{spec}$ (eq.~\ref{R_spec})
has a higher contribution are strongly suppressed under a white sky
distribution. Also note that by definition of the $D$ model, the
BHR simply equals the corresponding value of $A$.

The BHR values obtained using the parameters $\mathcal{O}_{DS}=\left\{ A_{min},n_{PTFE},\lambda=5\,\text{m}\right\} $
and $\mathcal{O}_{D}=\left\{ A_{min},\lambda=5\,\text{m}\right\} $,
shown in table~\ref{tab:results}, represent for each PTFE sample
the lowest reflectance that can be taken from this work at a $95\text{\%}$
confidence level. The relevance of these values relates directly to
the performance of LXe detectors using PTFE reflectors as they define
a lower limit for the detection threshold in terms of light collection.

It is worth mentioning that relative reflectance measurements as a
function of the PTFE thickness were also carried out in a different
setup for samples of both the 807NX and NXT85 materials immersed in
LXe \cite{Haefner:2016aa}. The obtained results for the scintillation
light of LXe show no appreciable reduction in the reflectance from
$9.5\,\text{mm}$ down to $1\,\text{mm}$ wall thickness. They also
provide the indication that the reflectance of NXT85 is slightly larger
than that of 807NX, in agreement with the results obtained in this
work for the absolute reflectance of those PTFE for the xenon scintillation
light.

\section{Conclusions\label{subsec:Conclusions}}

The absolute reflectance of three PTFE samples immersed in LXe was
measured for the xenon scintillation light ($\lambda=178\,\text{nm}$).
To the best of our knowledge, these are the first measurements of
the PTFE reflectance immersed in LXe to be performed using a dedicated
setup optimized for that purpose.

The experimental method used proved to be sensitive to small differences,
$\mathcal{O}$(0.1\%), in the absolute reflectance of the PTFE samples.
The obtained results confirm the higher reflectance of PTFE at $\lambda=178\,\text{nm}$
when immersed in LXe when compared with extrapolations from existing
measurements performed in gas and at room temperature. This is in
agreement with previous observations in large and medium size detectors
built for other purposes. Furthermore, the results show that very
high reflectances $(>97\text{\%})$ can be attained for xenon scintillation
light when using PTFE immersed in LXe. The results also support the
existence of a specular reflection component, which is usually not
considered in the simulation of LXe detectors based on xenon scintillation
and using PTFE as a light reflector. The addition of this specular
component may be very important (depending on the geometry) for the
correct reconstruction of events near the PTFE walls in such detectors
(e.g. LZ, LUX, XENON). The ability to choose and characterize the
PTFE reflectance early in the design stage may play a crucial role
in the optimization of future LXe detectors, especially in what concerns
their detection threshold and sensitivity.

\section*{Acknowledgments}

We would like to acknowledge the LZ and LUX collaborations for providing
the PTFE materials and for many useful discussions. This work was
supported by funding from Fundação para a Ciência e a Tecnologia (FCT)
through the Project-Grant No. PTDC/FIS-NUC/1525/2014.

\bibliographystyle{IEEEtran}
\bibliography{refs}

\end{document}